\documentstyle[12pt]{article}
\global\parskip 6pt

\normalsize

\let\oldtheequation=\theequation
\def\doteqs#1{\setcounter{equation}{0}
            \def\theequation{{#1}.\oldtheequation}}

\newcounter{sxn}
\def\sx#1{\addtocounter{sxn}{1} \bigskip\medskip \goodbreak
\noindent{\large\bf
\centerline{\thesxn.~~#1}} \nobreak \medskip}
\def\sxn#1{\sx{#1} \doteqs{\thesxn}}

\newcounter{axn}

\def\br{\begin{thebibliography}{abc}}
\def\er{\end{thebibliography}}

\def\be{\begin{equation}}
\def\ee{\end{equation}}
\def\bea{\begin{eqnarray}}
\def\eea{\end{eqnarray}}

\begin{document}
\begin{flushright}
\hfill{SINP-TNP/03-03}\\
\end{flushright}
\vspace*{1cm}
\thispagestyle{empty}
\centerline{\large\bf Hidden Degeneracy in the Brick Wall Model of Black
Holes}
\bigskip
\begin{center}
Kumar S. Gupta\footnote{Email: gupta@theory.saha.ernet.in}\\
{\em Theory Division\\
Saha Institute of Nuclear Physics\\
1/AF Bidhannagar\\
Calcutta - 700064, India}\\
\vspace*{.5cm}
Siddhartha Sen\footnote{Email: sen@maths.tcd.ie}\\
{\em School of Mathematics\\
Trinity College\\
Dublin , Ireland}\\
\vspace*{.1cm}
and\\
\vspace*{.1cm}
{\em Department of Theoretical Physics}\\
{\em Indian Association for the Cultivation of Science}\\
{\em Calcutta - 700032, India}\\
\end{center}
\vskip.5cm

\begin{abstract}

	Quantum field theory in the near-horizon region of a black hole
predicts the existence of an infinite number of degenerate modes. Such a
degeneracy is regulated in the brick wall model by the introduction of a
short distance cutoff. In this Letter we show that states of the brick
wall model with non zero energy admit a further degeneracy for any given
finite value of the cutoff. The black hole entropy is calculated within the
brick wall model taking this degeneracy into account. Modes with complex
frequencies however do not exhibit such a degeneracy.

\end{abstract}
\vspace*{.3cm}
\begin{center}
February 2003
\end{center}
\vspace*{1.0cm}
\newpage

\sxn{Introduction}

	The event horizon plays a central role in various aspects of black
hole physics. A simple illustration of this role of the horizon is
provided by the brick wall model proposed by 't Hooft \cite{hooft1,hooft2}.
This model describes the low energy dynamics of a scalar field the in the
background of a Schwarzschild black hole. The density of states near the
black hole horizon is however divergent, due to the presence of an infinite
number of degenerate modes in the near-horizon region
\cite{hooft1,hooft2,suss,hooft3}. Such a divergence is controlled in the
brick wall model through the introduction of a short distance cutoff (i.e. a
brick wall). For every finite value of the cutoff, this model provides a well
defined expression for the density of states and entropy. The entropy of the
black hole as determined from this model arises essentially from the
near-horizon features of the system.
 
	The above discussion suggests that the ambiguity in determining
the density of states in the brick wall model is completely removed by the
short distance cutoff. In this Letter we show that even with
such a finite cutoff, there is still an additional degeneracy 
present in the system. This degeneracy is revealed by a  study of
the brick wall model in the near-horizon region of the black hole. For a
given value of the cutoff, the allowed quantum states of this model are
labelled by energy and other quantum numbers. We shall show that for any
given finite value of the cutoff, every such allowed quantum state with
non-zero energy has infinite degeneracy.

	The infinite degeneracy found is expected to contribute to the black
hole entropy. In order to analyze this contribution, we show that  
a dimensionless parameter can be introduced to regulate the degeneracy. The
resulting expression for the black hole entropy is found to depend on the
ratio of this parameter to the brick wall cutoff. The assumption of
finiteness of this parameter is found to be consistent within the framework
of the brick wall model.

	Our analysis can be extended to the case of complex energy values.
Such states are analogous to the quasinormal modes of the Schwarzschild
black hole \cite{chandra}. Analysis of the equation satisfied by these
modes shows that the degeneracy disappears due to the presence of complex
frequencies.

	This Letter is organized as follows. In Section 2 we show how the
degeneracy appears in the brick wall model. Similar analysis
for modes with complex frequencies is performed in Section 3 and it is shown
that such modes admit no degeneracy. In Section 4 we discuss how the density
of states and black hole entropy are modified in presence of the new
degeneracy. Section 5 concludes this Letter with some discussions.


\sxn{Hidden Degeneracy in the Brick Wall Model}

The brick wall model \cite{hooft1,hooft2} describes the low energy quantum
dynamics of a scalar field $\phi$ in the background of a massive
Schwarzschild black hole of finite mass $M$. For simplicity we shall take the
scalar field to be massless. In the spherical polar coordinates coordinates
$(r, \theta, \phi)$, the field theory description is assumed to be valid in
the region $ r > R + h $ where $R = 2 M$ is the radius of the horizon and
the brick wall cutoff $h (> 0) $ is small compared to $R$. The field $\phi$
is assumed to satisfy the boundary condition
\be
\phi (r= R + h, \theta, \phi, t ) = 0.
\ee
In addition, the whole system is assumed to be in a box of radius $L$ which 
provides an infrared cutoff with the associated boundary condition
\be
\phi (r, \theta, \phi, t ) = 0 ~~~~{\mathrm {for}}~ r\geq L. 
\ee

	In the background of a Schwarzschild black hole, the 
field equation for modes with angular momentum $l$ and energy $E$ is
given by \cite{hooft2}
\be
\left ( 1 - \frac{2 M}{r} \right )^{-1} E^2 \phi + \frac{1}{r^2}
\frac{\partial}{\partial r} \left ( r (r - 2M) \frac{\partial \phi}
{\partial r} \right )
- \frac{l (l+1)}{r^2}\phi = 0.
\ee
In order to study the near horizon properties of the brick wall model, it is
useful to introduce a new coordinate $x \equiv r - R $. In the near-horizon
region, i.e. when $x \ll R$, Eqn. (3) can be written as
\be
\frac{\partial^2 \phi}{\partial x^2} + \frac{1}{x} \frac{\partial \phi}
{\partial x} + \frac{R^2 E^2}{x^2}  \phi - \frac{l (l+1) }{Rx} \phi = 0.
\ee 
In terms of a new field $\chi$ defined by
\be
\chi = \sqrt{x} \phi,
\ee
Eqn. (2.4) can be written as 
\be
\frac{\partial^2 \chi}{\partial x^2} + \frac{1}{x^2}
\left [ \frac{1}{4} + R^2 E^2 \right ] \chi - \frac{l (l+1) }{Rx} \chi
= 0.
\ee 
For small values of $x$, the angular term in the field Eqn. (2.6) can be 
ignored as has been noted elsewhere as well \cite{hooft3}. 
Thus the KG equation in the near-horizon region can be
written as
\be
\frac{\partial^2 \chi}{\partial x^2} + \frac{1}{x^2} 
\left [ \frac{1}{4} + R^2 E^2 \right ] \chi = 0.
\ee 

The field equation (2.3) has been analyzed by 't Hooft using the WKB method
\cite{hooft1,hooft2}. Such an analysis with the brick wall boundary
conditions is assumed to provide a discrete set of eigenvalues. Let us
consider Eqn. (2.7) with  such a generic eigenvalue $E$, the latter being
now treated as a parameter already determined by the WKB analysis. We can
think of as Eqn. (2.7) as a Schrodinger equation in presence of a singular
potential \cite{landau,accu} with $\chi$ now being interpreted as the zero
eigenvalue solution of such an equation. The central question that we address
here can now be stated as follows: for the allowed values of the parameter
$E$ and in the presence of the brick wall cutoff $h$, does the singular
quantum mechanical problem described by Eqn. (2.7) possess a unique solution or
is there any additional degeneracy which is perhaps not manifest?  In order
to address this issue, it is useful to start by considering a different
equation given by
\be
\frac{\partial^2 \chi}{\partial x^2} + \frac{1}{x^2}
\left [ \frac{1}{4} + R^2 E^2 \right ] \chi  = \eta^2 \chi.
\ee
In what follows, we shall interprete Eqn. (2.7) as the $\eta \rightarrow 0$
limit of Eqn. (2.8). We shall show below that in the limit when 
$\eta \rightarrow 0$, Eqn. (2.8) admits an infinite number of solutions even
in the presence of the brick wall cutoff. More precisely, we will see that
$\eta = 0$ is an accumulation point for the solutions of Eqn.
(2.8).

We now proceed to solve Eqn. (2.8) with the brick wall boundary conditions.
Eqn. (2.8) can be related to the equation for modified Bessel functions with
imaginary order. The two linearly independent solutions $u_1$ and 
$u_2$ of Eqn. (2.8) can be taken as \cite{accu}
\bea
u_1 &=& \sqrt{x} I_{-i\lambda} (\eta x),\\
u_2 &=& \sqrt{x} I_{i \lambda} (\eta x),
\eea
where $\lambda = RE $ and $I$ denotes the modified Bessel function
\cite{abr}.
The general solution of Eqn. (2.8) which vanishes at $x = h $ can thus be
written as
\be
\chi (x) = A \sqrt{x} \left [ I_{i \lambda} (\eta x) I_{-i \lambda} 
(\eta h) - 
I_{-i \lambda} (\eta x) I_{i \lambda} (\eta h)
\right ],
\ee 	
where $A$ is a constant. When $r \rightarrow L$, the coordinate $x 
\rightarrow (L - R)$. In this limit and under the condition that 
$(L - R) \gg h$, the leading 
divergent behaviour of $\chi (x)$ is given by
\be
\chi (x) =  \frac{A {\rm e}^{\eta x}}
{\sqrt{2 \pi \eta}}
\left [ I_{- i \lambda} (\eta h) - I_{i \lambda} (\eta h) 
\right ].
\ee
For $x \geq (L - R)$, Eqn. (2.2) requires that 
$\chi (x)$ must vanish. From Eqn. (2.12) we see that this is possible only if 
\be 
I_{ i \lambda} (\eta h) = I_{- i \lambda} (\eta h).
\ee
Let us now recall that we are interested in the situation where $h
\rightarrow 0$, i.e. where the brick wall is very close to the horizon. When
$h \rightarrow 0$, $I_{ i \lambda} (\eta h)$ behaves as
\cite{abr}
\be
I_{ i \lambda} (\eta h) \rightarrow \frac{(\eta h)^{i \lambda}}
{2^{i \lambda}} \frac{1}{\Gamma{(1 + i \lambda)}}.
\ee
Using Eqns. (2.13) and (2.14) we see that the eigenvalues of Eqn. (2.8) are 
given by
\be
\eta_p^2 = \frac{1}{h^2} 
{\mathrm exp} \left ( 2{\mathrm ln}2 + \frac{2 \theta
(\lambda)}{\lambda} + 2p \frac{\pi}{\lambda} \right  ),
\ee
where $p \in Z $ is an arbitrary integer and the function 
$\theta (\lambda)$ is the
argument of the Gamma function $\Gamma(1 + i \lambda)$. 
As mentioned before, we are 
interested only in the zero eigenvalue solutions of Eqn. (2.8). 
From Eqn. (2.15) we see that $\eta_p \rightarrow 0$ when $p \rightarrow 
- \infty$, or equivalently 
when $p = -n$ where $n$ is an arbitrarily large positive integer. 
In that case, Eqn. (2.15) can be written as 
\be
\eta_n^2 = \frac{1}{h^2} 
{\mathrm exp} \left ( - 2n \frac{\pi}{\lambda} \right  ).
\ee
From Eqn. (2.16) we
see that as $n \rightarrow  \infty $, there are an infinite number of zero
modes solutions of Eqn. (2.8). In other words, the zero is an accumulation
point for the eigenvalues of Eqn. (2.8).

We shall now point out several implications of the above analysis
:
\vskip 1 mm
\noindent
1) The energy eigenvalue $E$ obtained by the WKB analysis of Eqn. (2.3)   
serves as a parameter in the above analysis. The degeneracy obtained above
exists for every real value of $E$ except for $E=0$. In the latter case we
have $\lambda = 0$ and Eqn. (2.13) reduces trivially to an identity.\\
2) The infinite number of zero eigenvalue states of Eqn. (2.8) exist for
every finite value of the brick wall cutoff parameter $h$. This degeneracy
is therefore not removed by the brick wall cutoff. 

The above analysis thus shows that for any given real value of
$E \neq 0$ and for any finite value of the brick wall cutoff parameter $h$, 
there exist an infinite number of zero eigenvalue solutions of Eqn (2.8),
which can be interpreted as solutions of Eqn. (2.7). It is in this sense
that we claim about the existence of the additional degeneracy in the
near-horizon region as described by the brick wall model.
Finally it may be noted that the above result would continue to hold for
massive fields as well as the field equations become independent of the mass
term in the hear-horizon region of the black hole \cite{hooft3}.

\sxn{Analysis of Modes with Complex Frequencies}

	In this Section we shall analyze Eqn. (2.7) for the case of complex
frequencies. For the purpose of this Section, we use the notation $\Omega$
instead of $E$. Eqn. (2.7) is therefore replaced by
\be
\frac{\partial^2 \chi}{\partial x^2} + \frac{1}{x^2}
\left [ \frac{1}{4} + R^2 \Omega^2 \right ] \chi = 0,
\ee 
where $\Omega \equiv \omega_1 + i \omega_2 $ is taken to be complex. It is
tempting to identify $\chi$ in Eqn. (3.1) with the quasinormal mode
amplitude for the Schwarzschild black hole. Indeed, the equation for the
quasinormal modes for the Schwarzschild case \cite{ferr} reduces to Eqn.
(3.1) in the near-horizon region. There are however certain differences in
the boundary conditions obeyed by the quasinormal modes \cite{chandra,ferr}
to the ones that are used in the brick wall model. In what follows we shall
analyze Eqn. (3.1) with the brick wall boundary conditions.

As described in Section 2, we interprete Eqn. (3.1) as the $\gamma
\rightarrow 0$ limit of the equation
\be
\frac{\partial^2 \chi}{\partial x^2} + \frac{1}{x^2}
\left [ \frac{1}{4} + R^2 \Omega^2 \right ] \chi = \gamma^2 \chi.
\ee 
As before, we are interested only in solutions of Eqn. (3.2) 
with $\gamma \rightarrow 0$. The general solution of
Eqn. (3.2) which vanishes at $x = h$ is given by
\be
\chi (x) = B \sqrt{x} \left [ I_{i \mu} (\gamma x) I_{- i\mu}    
(\gamma h) - I_{- i\mu} (\gamma x) I_{ i\mu} (\gamma h)
\right ],
\ee 
where $\mu = R \Omega$. When $ x \geq (L-R)$ and $(L-R) \gg h$, 
the wavefunction is Eqn. (3.3) can vanish only if 
\be
I_{i \mu} (\gamma h) = I_{ -i \mu} (\gamma h).
\ee
In the limit when $h \rightarrow 0$, Eqn (3.4) gives
\be
(\gamma h )^{2 i\mu}~ 2^{2i\mu} = \frac{\Gamma (1 + i\mu)}{\Gamma (1 -
i\mu)}.
\ee
Since $\Omega = \omega_1 + i \omega_2$,
the gamma functions in Eqn. (3.5) can be expressed in the polar form as
\bea
\Gamma (1 + i\mu) = \Gamma (1 - R \omega_1 + i R \omega_1) & \equiv & 
\xi_1 {\rm e}^{i \theta_1} \\
\Gamma (1 - \mu) = \Gamma (1 - R \omega_2 - i R \omega_1) & \equiv & 
\xi_2 {\rm e}^{i \theta_2}.
\eea 
Using Eqns. (3.6) and (3.7), the imaginary and real parts of Eqn. (3.5) can
be written as 
\bea
2 R \omega_1 \left ( {\mathrm ln} \gamma h + {\mathrm ln} 2 \right )
&=& (\theta_1 - \theta_2) + 2 p \pi \\
-2 R \omega_2 \left ( {\mathrm ln} \gamma h + {\mathrm ln} 2 \right )  
&=& {\mathrm ln} \frac{\xi_1}{\xi_2} ,
\eea
where $p \in Z $ is an arbitrary integer. However, the zero eigenvalue
solutions of Eqn. (3.2) are obtained only when $p = -n$ where $n$ is an
arbitrarily large positive integer. In that case Eqn. (3.8) can be written as  
\be
{\mathrm ln} \gamma_n h = - \frac{  n \pi}{R \omega_1}.
\ee
Using Eqns. (3.9) and (3.10) we get
\be
\left (\frac{\omega_1}{2 \pi \omega_2} \right )
{\mathrm ln} \frac{\xi_1}{\xi_2} = n.
\ee
Since $n$ is an arbitrarily large, for fixed values 
of $\omega_1$ and $\omega_2$ Eqn. (3.11) has no solution. We can therefore
conclude that there is no additional degeneracy for the case of the
modes with complex frequencies.

\sxn{Black Hole Entropy}

As discussed in Section 2, the solutions of Eqn. (2.8) for any non-zero
value of $E$ admit an infinite degeneracy due to the near-horizon effects. 
Let us also recall that the black hole entropy found in the brick wall model
arises essentially from the near-horizon features of the model. It is
therefore expected that the degeneracy found in Section 2 would contribute
to the black hole entropy, as they are both related to the physics in the
near-horizon region. Below we shall discuss the effect of this degeneracy on
the density of states and black hole entropy.

We shall first discus how the degeneracy found in Section 2 
affects the density of states. For this purpose, 
consider two eigenvalues $\eta_{n_1}$ and $\eta_{n_2}$ of the Eqn. (2.8)
where the two positive integers $n_1, n_2 \rightarrow \infty$. For 
definiteness let us assume that $n_2 > n_1$. From Eqn. (2.16) we get
\be
\Delta n \equiv n_2 - n_1 = 
\frac{\lambda}{\pi} {\mathrm ln} \left ( \frac{\eta_{n_1}}
{\eta_{n_2}} \right ) ~~~~~~~(\lambda = RE).
\ee
The quantity $\Delta n$ provides an expression for the degeneracy 
of the quantum state with energy $E$. The logarithm on the r.h.s. of Eqn.
(4.1) contains the ratio of two numbers both of which are tending towards
zero. We shall assume that the logarithm of this ratio tends to a positive 
constant $c$ as $n_1, n_2 \rightarrow \infty$. This is an ad hoc assumption 
which is nevertheless consistent within the framework of this model. Eqn.
(4.1) can now be written as
\be
\Delta n = \frac{c \lambda}{\pi} = \frac{c R E}{\pi}.
\ee 
The above expression vanishes for $E=0$, which is consistent with the
observation made at the end of Section 2. The density of states calculated
in the brick wall model would now have to be multiplied with this degeneracy
factor. The corresponding expression of the free energy $F$ obtained from
the brick wall model \cite{hooft1,hooft2} is now modified to read as 
\be
\pi \beta F = - \frac{2}{3} \frac{(2M)^4}{h} \int_0^{\infty}
\frac{(\Delta n) E^3 dE}{e^{\beta E} -1}.
\ee
The corresponding expression of the black hole entropy is now obtained as 
\be
S = \frac{80 }{\pi^3 } \frac{c}{h} \frac{(2M)^4}{\beta^3}.
\ee
On the other hand, 
the application of quantum mechanical scattering theory to this 
system gives the black hole entropy as \cite{hooft1,hooft2}
\be
S = \frac{4 \pi M^2}{G},
\ee
where $G$ is the Newton's constant in units chosen in Ref. \cite{hooft2}.
Comparing Eqns. (4.4) with (4.5) we get
\be
h = \frac{320~ G~c}{\pi^7 M}.
\ee 
We therefore see within the framework presented above, the main effect of
the additional degeneracy is to modify the expression of the brick wall
cutoff parameter $h$ (compare with Eqn. (8.19) of Ref. \cite{hooft2} with
$N=1$). Finally, the requirement that physical distance of
the brick wall from the horizon be of the order of Planck length
\cite{hooft2}, leads to a finite value of the constant $c$ which
is consistent with assumption made above.

\sxn{Conclusion}

	In this Letter we have shown that there is a hidden degeneracy in
the brick wall model in the near-horizon region of the black hole. This
degeneracy has been found by the analysis of the zero eigenvalue solutions
of Eqn. (2.8) which can be identified with the solutions of Eqn. (2.7).
Usually, any reasonable quantum field theory is expected to predict an
infinite number of modes in the near-horizon region of a black hole
\cite{hooft1,hooft2,suss,hooft3}. The divergence from these modes is handled
in the brick wall model by the introduction of a short distance cutoff. The
degeneracy that we have discussed is different from the usual one in the
sense that it is not removed by the brick wall cutoff. We have in fact shown
that for any finite value of the brick wall cutoff, every quantum state of
the system with non-zero energy still admits an infinite degeneracy.

	The additional degeneracy discussed above is essentially a
near-horizon feature and is thus expected to affect the black hole entropy.
We have estimated the effect of this degeneracy in terms of a parameter $c$,
which has been assumed to be finite. The contribution of this degeneracy to
the density of states is finite and non-zero except for $E=0$. The 
corresponding expression for the free energy and entropy as calculated 
within the brick wall depends on the ratio $\frac{c}{h}$.
The requirement that the physical distance of the brick wall from the
horizon is of the order of Planck length leads to a finite value of $c$. The
assumption of finiteness of $c$ is thus consistent within the framework of
the brick wall model.

	We have also analyzed the system with complex frequencies. Such a
situation arises in the description of quasinormal modes of Schwarzschild
black holes. It has been shown that for finite values of the brick wall
cutoff there is no further degeneracy in this case.

	It has been argued in Ref. \cite{hooft3} that for a wide variety of
quantum fields including massive ones, the field equation in the near
horizon region looks essentially the same. The analysis presented in this
paper is thus expected to hold for all such fields and also for a variety of
black hole metrics with similar near-horizon structures. It is plausible
that the degeneracy discussed in this Letter is related to the conformal
symmetry which is present in the near-horizon region of a wide class of
black holes \cite{carlip,kumar,moretti}.


\bibliographystyle{unsrt}

\begin{thebibliography}{99}
\bibitem{hooft1} G. 't Hooft, Nucl. Phys. {\bf B256} (1985) 727.
\bibitem{hooft2} G. 't Hooft, Int. Jour. Mod. Phys {\bf A11} (1996) 4623.
\bibitem{suss} L. Susskind and J. Uglum, Phys. Rev. {\bf D50} (1994) 2700.
\bibitem{hooft3} G. 't Hooft, {\it The Holographic Principle}, 
Opening lecture at the International School of Subnuclear Physics, Erice, 1999, 
hep-th/0003004.
\bibitem{chandra} S. Chandrasekhar, {\it The Mathematical Theory of Black
Holes}, Clarendon, Oxford, 1983.
\bibitem{landau} L. D. Landau and L. M. Lifshitz, {\it Quantum Mechanics},
Pergamon Press, Oxford, 1958. 
\bibitem{accu} K. M. Case, Phys. Rev. {\bf 80}, 797 (1950); P. Morse and
H. Feshbach, {\it Methods of Theoretical Physics, Vol. 2}, McGrawHill, New
York, 1953.
\bibitem{abr} M. Abromowitz and I. A. Stegun, {\it Handbook of Mathematical 
Functions}, Dover, New York, 1970.
\bibitem{ferr} V. Ferrari and B. Mashhoon, Phys. Rev. {\bf D30} (1984) 295.
\bibitem{carlip} S. Carlip, Phys. Rev. Lett. {\bf 82} (1999) 2828; Class.
Quant. Grav. {\bf 16} (1999) 3327; Nucl. Phys. Proc. Suppl. {\bf 18} (2000) 
10; M.-I.Park, Nucl. Phys. {\bf B364} (2002) 339.
\bibitem{kumar} D. Birmingham, Kumar S. Gupta and Siddhartha Sen, Phys. Lett. 
{\bf B 505} (2001) 191.
\bibitem{moretti} V. Moretti and N. Pinamonti, Nucl. Phys. {\bf B647} (2002)
131.
\end{thebibliography}

\end{document}